\documentstyle[12pt]{article}
\def\QP{quasiparticle}
\def\QPs{quasiparticles}
\begin{document}

\title{Electrons in a strong magnetic field on a disk}
\author{M. Kasner\thanks{\noindent Present address: Dept.\ of Physics,
        Indiana University, Swain Hall West 117, Bloomington, IN 47405, USA}
        \ and W. Apel \\Physikalisch-Technische Bundesanstalt
        \\Bundesallee 100 \\D--38116 Braunschweig, Germany}
\date{}
\maketitle

\begin{abstract}
\noindent
The problem of interacting electrons moving under the influence of a strong
magnetic field in two dimensions on a finite disk is reconsidered.
First, the results of exact diagonalizations for up to $N=9$ electrons for
Coulomb as well as for a short--range interaction are used in the search for
a peculiar ground state corresponding to filling factor $1/3$.
Not for the Coulomb, but only for the short--range interaction, can the
$1/3$--state be safely identified amongst the spectra of various filling
factors
close to $1/3$.
Second, the propositions of the concept of quasiparticles, as used in the
hierarchical theory, are examined in view of the exact results for the disk
geometry.
Whereas the theory for the quasiholes is in complete accordance with the
spectra, for the quasielectrons, finite size corrections make an analysis
difficult.
For the quasielectron energy, an extrapolation to $N \rightarrow \infty$ is
given and compared with the corresponding extrapolations of three different
proposals for trial wave functions.
While the limiting value for the best trial wave function is very close to the
limit of the exact results, the behavior of the finite size corrections of the
exact energies and of the trial wave functions, respectively, is qualitatively
rather different.
\end{abstract}

\noindent
\begin{flushleft}  PACS-Numbers:  73.40.Hm, 73.20.Dx   \end{flushleft}

\newpage
\section{Introduction}
The peculiar transport properties of a two--dimensional electronic system in a
strong magnetic field, as seen in the integral quantum Hall effect (IQHE)
\cite{vKDP80} and the fractional quantum Hall effect (FQHE) \cite{TSG82}, are
just a part of the rich structure of the phase diagram in the electron density
vs.\ temperature plane \cite{KPvKHPT93,KLZ92}.
Much work was devoted in the last years to a study of the Wigner crystal phase
at low densities \cite{KPvKHPT93} and, recently, to the interesting properties
at filling factor $1/2$ \cite{HLR93,DSTPW93,WRPWP93}.
Here, we wish to reconsider the FQHE reporting work done on the few--particle
problem. \\
In both quantum Hall effects, there is a gap above the ground state at the
magic filling factors in the ordered system.
If a gap remains in the presence of disorder, a plateau in the Hall
conductivity
$\sigma _{xy}$ is to be expected in the experiments (around filling factors
$\nu =p/q$ in the FQHE ; $p, q$ -- relative primes, $q$ -- odd);
the longitudinal conductivity $\sigma _{xx}$ vanishes wherever $\sigma _{xy}$
is in a plateau region.
Despite the experimental similarity in the transport properties, the
microscopic
models considered for the IQHE and FQHE are quite different.
The IQHE is understood as a localization--delocalization transition, as shown
by
non--interacting electrons moving in a strong magnetic field under the
additional influence of a moderate random potential \cite{Pru90}.
There, the energy gap is due to the cyclotron energy.
In contrast, in the FQHE, even the starting point of a theoretical analysis,
namely interacting spinpolarized electrons in the lowest Landau level without
any disorder, is a non--tractable system itself because of the strong
correlation.
Here, even the existence of an energy gap in the many--particle spectra at the
magic filling factors has to be proven.  \\
An important step towards an understanding was taken by Laughlin
\cite{Lau83a,Lau83b,Lau90a} who showed the special character of the ground
state
at filling factors $1/q$ ($q$ -- odd) by presenting a trial wave function.
This success was partially based on conclusions drawn from few particle
calculations.
Laughlin's notion of quasiparticles in connection with the hierarchical theory
\cite{Halp83,Hald83} opened a way to explain the occurence of filling factors
with numerators different from one, e.\ g.\ $2/5$ .
However, the hierarchical theory is rather qualitative and allows all filling
factors with an odd denominator; thus, a further theoretical analysis becomes
necessary.
Moreover, a microscopic derivation of an effective quasiparticle Hamiltonian
which would start from the original electronic Hamiltonian is still missing. \\
A new and different point of view emerged in the work of Jain
\cite{Jai89a,Jai90a,Jai92}.
He circumvented the rather artificial construction of many particle states at
$\nu =p/q$ via quasiparticles which is done in the hierarchical scheme.
Instead, he suggested explicit electronic wave functions for all observed
filling factors, $\nu =n/(2np+1)$ ($n,p$ -- integers) and particle--hole
transformed states.
Jain's approach starts from $n$ completely filled Landau levels, then
multiplies
the wave function with a symmetric Jastrow factor of power $2p$ ("adds $2p$
flux
quanta per electron") in order to simulate the influence of the interaction
and,
finally, projects to the lowest Landau level.
Numerical calculations for a small number of particles \cite{DJ92b}, and also
the Fermi--liquid like behavior seen in the experiments \cite{DSTPW93,WRPWP93}
near $\nu =1/2$ corraborate Jain's approach.
Nevertheless, the reason for its success remains unclear if one considers the
quite different energy scales of non-interacting and interacting electrons,
respectively \cite{Mac92}. \\
Other approaches attack the problem with field theoretical methods
\cite{Gir90a,ZHK89a,Zha92}.
These attempts describe already known features of the theory such as the
Laughlin ground state, the nature of the quasiparticles and the collective
excitation spectrum \cite{GMP86a}. \\
Both, the trial wave function approach and the field theories consider only
points of the phase diagram of interacting electrons in the lowest Landau
level,
because the analysis focuses on definite filling factors.
It should be emphasized that the approaches considered so far have another
thing
in common: they do not depend very much on the specific form of the
electron--electron interaction.
On the one hand, this can be seen as expressing a universality of the FQHE.
On the other hand, one would like to check such an assertion.
A nice concept for a discussion of the interaction is
that of the pseudopotential
coefficients, i.\ e.\ the eigenenergies of the two-particle problem, introduced
by Haldane \cite{Hald90}.
In the lowest Landau level, the interaction is completely determined by these
coefficients.
The picture emerged, that the largest of these determines the FQHE ground
state,
not to be reached from the non--interacting ground state by perturbation
theory,
while the other coefficients can be included perturbatively and do not lead to
a
big difference.
Is this true? Below, we will show the numerical spectra for two different
interactions (i) the full Coulomb interaction and (ii) a pseudopotential
interaction, where only the first coefficient is non--zero. \\
At first glance, the choice of an appropriate geometry for studying such a
system seems to be only a matter of convenience. Three geometries were studied
in the past: the disk geometry with a symmetric gauge and open boundary
conditions \cite{Lau83b}, the torus geometry with periodic boundary conditions
based on the Landau gauge, cf.\ \cite{YHL83}, and the spherical geometry where
the electrons move in the constant magnetic field of a magnetic monopole in the
centre \cite{Hald83}. The analytical theory of Laughlin \cite{Lau90a} was
formulated on the disk and also numerical work was done for this geometry
\cite{Lau83b,GJ83,LYSY84,MH86}. Soon, the advantages of the spherical geometry
with an additional symmetry and without boundary attracted a lot of interest.
It became the favorable choice allowing exact diagonalizations for up to ten
electrons at filling factor $1/3$
\cite{FOC86} and the study of the hierarchical theory, because the
lower energy levels are well separated from the rest of the spectrum. \\
In this paper, we come back to the disk geometry for several good reasons:
-- From the experimental point of view, one should study a planar geometry,
particularly, if one wants to account for boundaries and contacts. \\
-- The topological differences between the different geometries can very well
change the results.  E.\ g.\ , the ground state at $\nu =1/q$ is nondegenerate
in the disk and spherical geometry, whereby in the torus geometry it is
$q$--fold degenerate \cite{Su84,Hald85,WN90}. \\
-- The Laughlin theory was originally formulated for the plane.
Then, in the current understanding of the FQHE, the plateaus of the Hall
conductivity in the vicinity of the magic filling factors are a result of the
disorder.
Up to now, calculations including disorder study the energy gap between
the ground state and the excited states, see e.\ g.\ \cite{RH85,MLGP86,Mie90a};
the calculation of the transport properties of a disordered and interacting
electronic system in a magnetic field is an outstanding problem.
Still, current theories of disorder use a planar geometry, see \cite{LR85}. \\
-- If one wants to verify theories stressing the importance of edge excitations
in the FQHE \cite{LW91}, one needs a geometry with an edge. \\
In this work, we present the results of a numerical study of the low--lying
eigenstates of the FQHE Hamiltonian at and in the vicinity of filling
factor $1/3$ for the disk geometry.
We compare the Coulomb with a pseudopotential interaction.
While on the sphere, the differences between both interactions are not so
marked, on the disk, the pseudopotential interaction plays a special role
and this will be shown below.
With our results, we evaluate properties of quasiparticles and thus check
the foundations of the hierarchical theory, by comparing the exact spectra for
small numbers of quasiparticles with energy expectation values of trial wave
functions. \\
The paper is organized in the following way: in Section 2, the model with the
interactions is introduced.
Section 3 describes shortly the numerical method used for the diagonalization.
In Section 4, we set out to identify the stable state in a system with a finite
number of electrons, and we compare the result with Laughlin's ground state
wave
function.
The notion of quasiparticles is discussed in general in Section 5.
Then, the results for the quasihole are discussed in Section 6, those for the
quasielectron in Section 7.
Here, we have extended our recent analysis \cite{KA93a} to a larger number of
electrons. Section 8 gives the summary.

\section{Model}
We consider electrons of charge $-e\, (e>0)$ moving in the x--y--plane under
the influence of a perpendicular, constant magnetic field
$\vec{B}=-B\bf\vec{e}_{z}$ ($B>0$).
They interact via an isotropic, translationally invariant interaction
$V(|\vec{r}-\vec{r'}|)$.
We study the disk geometry and then, the appropriate gauge is
the symmetric gauge $\vec{A}=\frac{B}{2}(y,-x,0)$. \\
The one--particle problem is equivalent to an isotropic two-dimensional
harmonic oscillator, whose Hamiltonian $H_{0}$ and angular momentum $L$ can
be expressed by two pairs of commuting Bose
operators $a,a^{\dagger}$ and $b,b^{\dagger}$, so that $H_{0}=\hbar
\omega_{c}
(b^{\dagger}b+\frac{1}{2})$ with $\omega _{c}=\frac{|eB|}{m}$ and $L=\hbar
(a^{\dagger}a - b^{\dagger}b)$.
Then, the one--particle basis is
\begin{equation}
|n,m> = [n!(n+m)!]^{-\frac{1}{2}}\: (b^{\dagger})^{n}\:
(a^{\dagger})^{n+m}\:|0>
\label{basis}
\end{equation}
with $ n=0,1,\ldots; m = -n,\ldots,0,\ldots$.
The energies $E_{n}=\hbar \omega _{c}(n+\frac{1}{2})$ are degenerate in each
of the Landau levels $n$ with respect to the angular momentum $m$.
Because the gap between adjacent Landau levels grows linearly with $B$,
whereas the coupling constant for a $1/r$ interaction increases only with the
square root of $B$, in the strong field limit (FQHE), the Landau level mixing
can be neglected and the Hilbert space of the one--particle basis (\ref{basis})
is restricted to $n=0$.
Analogously, if the g--factor in the Zeeman term is not too small, the
electrons are spin polarized (about the necessity to include the spin, see
\cite{BM91}).
The basis (\ref{basis}) becomes in real--space representation for the lowest
Landau level
\begin{equation}
\varphi _{m}(z)=[2\pi 2^{m}m!]^{-\frac{1}{2}} \: z^{m} \:
 e^{-\frac{1}{4} |z|^{2} } \:,
\label{llbasis}
\end{equation}\\
where $z=x+iy$ and all lengths are given in units of
$l_{c}=\sqrt{\hbar /|eB|}$. \\
We want to calculate the energy spectrum for a finite number $N$ of electrons
at a given filling factor. The "filling factor" $\nu '$, naively is the ratio
of the number of electrons to the number of available one--particle states.
In the disk geometry with background potential, this latter number is given by
the number $N_{\Phi }$ of flux quanta $\frac{h}{e}$ through the area of the
background.
For the neutralizing background potential, we take a superposition of all
one--particle states with an angular momentum from $m=0$ to $m_{max}$; this
leads in the thermodynamic limit to a homogeneous background charge density.
Then, outside the disk with the area $2\pi N_{\Phi } =2\pi(m_{max}+1)$,
the background charge density drops quickly to zero.
Thus, there are just $m_{max}+1$ one--particle states with a maximum
expectation value of the area not exceeding the area of the background.
Therefore, $\nu' = \frac{N}{N_{\Phi }} = \frac{N}{m_{max}+1}$.
For the numerical diagonalization, we restrict the single electron Hilbert
space
to only $h+1$ one--particle states with $m=0, 1, \ldots, h$.
The Hamiltonian to be diagonalized numerically then reads:
\begin{eqnarray}
H & = & \frac{1}{2}\sum\limits_{m_{1},m_{2},m_{3},m_{4}=0}^{h}
W_{m_{1},m_{2},m_{3},m_{4}} \: c^{\dagger}_{m_{1}}c^{\dagger}_{m_{2}}c_{m_{3}}
c_{m_{4}} \nonumber \\
& & - \nu^{'} \sum\limits_{m=0}^{h}\sum\limits_{m'=0}^{m_{max}}
W_{m,m',m',m} \: c^{\dagger}_{m}c_{m}
 + \frac{\nu'^{2}}{2} \sum\limits_{m,m'=0}^{m_{max}}
W_{m,m',m',m} \;,
\label{H}
\end{eqnarray}
where the kinetic energy term has been already dropped since it is a
constant for fixed $N$. The $c_{m}, c_{m}^{\dagger}$ are Fermi annihilation and
creation operators and the two--particle matrix element is
\begin{equation}
W_{m_{1},m_{2},m_{3},m_{4}} = \int \int \varphi _{m_{1}}^{*}(z)\varphi
_{m_{2}}^{*}(z')V(|z-z'|)\varphi _{m_{3}}(z')\varphi _{m_{4}}(z)\, d^{2}z
d^{2}z'.
\end{equation}
\\
In the following, we specify our two different choices for interaction
potential
$V(|z|)$ and confinement.
The complete model (\ref{H}) for the case of a Coulomb interaction with
background, the Coulomb model (CM), is the first one.
The second is a short--range interaction, the short--range model (SRM).
In contrast to the first case, here is no neutralizing background and the
confinement is due to the truncation of the one--particle basis at $h=m_{max}$.
In this case, we will use a slightly different definition for the filling
factor, $\nu = \frac{N-1}{N_{\Phi }-1} = \frac{N-1}{m_{max}} $,
because of reasons which will become clear below.
The difference between $\nu$ and $\nu '$ disappears in the thermodynamic limit.
Then, in the case of the SRM, the Hamiltonian reduces to the first term in
(\ref{H}) with $h=m_{max}$. \\
In first quantization, the first term of $H$ can be decomposed as follows
\begin{equation}
H^{1st}  =   \sum_{i>j=1}^{N}\sum_{m=0}^{\infty} V_{m}
\hat{P}^{(m)}_{ij} \:.
\label{H1st}
\end{equation}
$\hat{P}^{(m)}_{ij}$ is the projector which annihilates in a
two--particle wave function all components except the one with relative angular
momentum $m$.
This decomposition in angular momenta was originally introduced by Haldane
for the sphere. The $V_{m}$ are the eigenvalues of
the two--particle problem and are called pseudopotential coefficients.
Because of the fermionic nature of the problem, only odd $m$ ($m=1,3,\ldots
$) contribute to the sum in (\ref{H1st}).
The $W_{m_{1},m_{2},m_{3},m_{4}}$ are expressed by the $V_{m}$ as follows :
\begin{eqnarray}
W_{m_{1},m_{2},m_{3},m_{4}} = \delta _{m_{1}+m_{2},m_{3}+m_{4}} \prod
_{i=1}^{4}(2^{m_{i}}m_{i}!)^{-\frac{1}{2}}
\sum _{m=0}^{m_{1}+m_{2}}(m_{1}+m_{2})! m!
V_{m} \times \nonumber \\ \times \sum_{\lambda = -\infty}^{\infty}
{m_{1} \choose \lambda } {m_{2} \choose m - \lambda } (-1)^{(m-\lambda) }
\sum_{\lambda' = -\infty}^{\infty}
{m_{3} \choose \lambda' } {m_{4} \choose m - \lambda' }
(-1)^{\lambda' }.
\label{W}
\end{eqnarray}
\\
The Coulomb model is specified by the pseudopotential coefficients
\begin{equation}
V_{m}^{CM}=\frac{\sqrt{\pi }(2m)!}{2^{2m+1}(m!)^{2}} \;.
\end{equation}
Here and below, all energies are measured in units of the coupling constant
$\frac{e^{2}}{4\pi \varepsilon l_{c}}$ of a Coulomb potential.
The SRM is specified by the coefficients
\begin{eqnarray}
V_{1}^{SRM} =  \frac{\sqrt{\pi}}{4}, \; \;  \;   \;  V_{m>1}^{SRM} =  0 .
\label{SRM}
\end{eqnarray}
Here, only electrons with relative angular momentum one interact and repel each
other.
An interaction in real space, which yields $V^{SRM}_{m}$ as pseudopotential
coefficients in the lowest Landau level, was introduced by Pokrovsky and
Talapov \cite{PT85}:
\begin{equation}
H^{1st}_{SRM} = V^{SRM}_{1}
\sum_{i>j=1}^{N} \nabla ^{2} \delta ^{2}(\vec{r_{i}}-\vec{r_{j}}) \:,
\end{equation}
hence the name short range model.
Why should one study a SRM ?
The ground state of (\ref{H}) at the magic filling factors $\nu$, where the
experiment shows the FQHE, cannot be reached by a simple perturbation theory
from non--interacting electrons.
If, on the other hand, the non--perturbative ground state of a SRM is already
very close to the true ground state of (\ref{H}),
then the higher coefficients ($V_{m > 1}$ in our case) can be switched on
perturbatively and do not change the essence of the results.
In this case, a SRM should be considered as a generic model for the FQHE.
Indeed, the Laughlin wave function is the exact ground state for $\nu =1/q$ for
a SRM with $V_{m \geq q} = 0$ \cite{Hald83,PT85,TK85}.
Particularly on the sphere, this reasoning is very successful.
In addition, Haldane showed for $\nu =1/3$ that if the $V_{m \geq 3}$ become
too
large compared with $V_{1}$, then the preference of the Laughlin wave function
as the ground state wave function is destroyed \cite{Hald90}.\\
For future reference, we give the translation operator.
The generator of the magnetic translations is $\vec{t}=\vec{p}-e\vec{A} $ and
commutes with the Hamiltonian for an infinite system.
For $N$ electrons, the magnetic translation is a product of unitary
one--particle operators $\hat T^{(i)}_{\xi }$
\begin{equation}
\hat T_{\xi }= \prod_{i=1}^{N} \hat T^{(i)}_{\xi }=\exp(i \sum_{i=1}^{N}
\vec{\xi }
\vec{t_{i}})=\exp(-\frac{1}{\sqrt{2}} \sum_{i=1}^{N}( a_{i}^{+}\xi ^{*}-a_{i}
\xi )) \:,
\end{equation}
where we have used again the bosonic operators $a,a^{+}$.\\
The action of the operator $\hat T^{(1)}_{\xi }$ on the basis (\ref{llbasis})
yields
\begin{equation}
\hat T^{(1)}_{\xi }\varphi _{m}(z_{1})
   =[2\pi 2^{m}m!]^{-\frac{1}{2}} \: (z_{1}-\xi)^{m} \:
 e^{-\frac{1}{4} |z_{1}|^{2} +\frac{1}{2}z_{1}\xi ^{*} -\frac{1}{4} |\xi|^{2} }
\:.
\label{T1}
\end{equation}
Next, the factors $(z_{1}-\xi )^{m}$ and $e^{\frac{1}{2}z_{1}\xi ^{*}}$
on
the r.\ h.\ s.\ of (\ref{T1}) can be expanded with respect to $\xi , \xi
^{*}$.
Thus, the transformed function ( a "ring" of radius $\sqrt{2(m+1)}$ around
$\xi $) is a superposition of all the degenerate eigenfunctions of the lowest
Landau level with angular momenta reaching up to infinity.
For an electronic system of restricted size, one needs to truncate the angular
momentum at $m_{max}$.

\section{Numerical treatment}
We want to study the energy spectrum, particularly the ground state and the
lowest excited states, of $H$ at a fixed filling factor.
Then, the electron number $N$ determines the area of the system expressed by
the degeneracy $N_{\Phi }$ or by $m_{max}$.
The dimension of the fermionic many--particle Hilbert space
$ {N_{\Phi}  \choose N}= {m_{max}+1 \choose N} $ grows dramatically with
increasing $N$. However, also for a finite system, the total angular momentum
$ M=\sum_{m=0}^{\infty} m c^{\dagger}_{m}c_{m}$ is conserved.
Thus, the diagonalization can be performed in a Hilbert sub--space with fixed
$M$, i.\ e.\ , in the "block $M$".
The allowed values of $M$ are $M_{min}=N(N-1)/2\leq M \leq M_{max}=
N(2m_{max}-N+1)/2$. \\
For the diagonalization of (\ref{H}) in a block $M$, we construct the many--
particle basis in terms of the Slater determinants of the one--particle states.
Then, the sum of the $m_{i}$ ($1 \leq i \leq N, 0 \leq m_{i} \leq m_{max}$) is
$M$ and all one--particle angular momenta have to be different (fermionic
case).
The dimensions of the blocks $M$ are symmetric between $M_{min}$ and $M_{max}$.
The matrix with $M$ being in the middle has the largest dimension and its
size limits the feasibility of the diagonalization.
The number of these many--particle basis states with fixed $M$,
$g_{F}(N_{\Phi },N,M)$, can be determined either numerically by a recursive
formula or from the following generating function \cite{Hel93}
\begin{equation}
x^{N(N-1)/2} \prod_{l=1}^{N}\frac{1-x^{(N_{\Phi }-(l-1))}}{1-x^{l}}
 = \sum_{M=M_{min}}^{M_{max}} g_{F}(N_{\Phi },N,M) x^{M}.
\label{gF}
\end{equation}
As an example, the largest Hilbert sub--space for $\nu =1/3$ and
10 electrons has $M=135$ and the dimension 246 448.
The maximum dimension is the same as that in the spherical geometry for the
same
number of electrons at the same filling factor, cf.\ \cite{FOC86}. \\
The $W_{m_{1},m_{2},m_{3},m_{4}}$ in (\ref{H}), determined by the
pseudopotential coefficients via (\ref{W}), are used to calculate numerically
the matrix elements between the many--particle states;
the diagonal elements are sums of the difference between the direct and the
exchange terms, the off--diagonal elements are only single differences.
With increasing $N$, more and more of the matrix elements are zero because of
the angular momentum conservation and, therefore, the matrix becomes sparse.
At filling factor $\nu =1/3$, for $N=6$, the matrix $M=45$ (dimension 338)
contains $6 108$ non--zero off--diagonal matrix elements out of $56 953$,
whereas for $N=9$ and $M=108$ (matrix dimension $45 207$), there are
$3 016 844$ non--zero matrix elements of $1 021 813 821$, i.\ e.\
only $0.3 \%$ off--diagonal elements are non--zero.
The number of non--zero off--diagonal elements grows approximately as the
matrix dimension to the power 1.25.
We have checked that there is no further (accidental) decomposition of a
block $M$ into subblocks as long as we use the above basis of Slater
determinants. \\
For the diagonalization, we apply two methods. For $N \leq 6$, we calculate
all energy levels by combining the Householder and the QR method.
For electron numbers $N \geq 7$ when the matrices become quite large and
sparsely occupied, we use a Lanczos method in which we restrict the search
for energy levels to the low--energy region. \\
In this work, we concentrate on the filling factor 1/3 and calculate the exact
energy levels. We do not study overlaps between eigenfunctions and trial wave
functions because our conclusions can be drawn already on the basis of quantum
numbers and expectation values.
In the spherical geometry, in contrast, one needs also the eigenfunctions for
a determination of the angular momentum $\vec{L}^{2}$.

\section{The stable state}
In the following, we describe the results of our numerical calculations for
a filling factor of precisely $1/3$, where the experiment shows the FQHE.
We study first the CM, then the SRM. \\
How can one identify {\em in a series of spectra for various $m_{max}$} at
fixed $N$ the one at filling factor $1/q$?
An important hint is given by the trial wave function for filling factor $1/q$,
the Laughlin wave function \cite{Lau83b}:
\begin{equation}
\Psi _{\frac{1}{q}}(z_{1},\ldots,z_{N})= \prod_{i>j=1}^{N}(z_{i}-
z_{j})^{q}\exp{\left[-\frac{1}{4}\sum_{i=1}^{N}|z_{i}|^{2}\right]}.
\label{Laughlin}
\end{equation}
Its total angular momentum is $M=qN(N-1)/2 \equiv M_{N}(q)$ ; we will
abbreviate $M_{N}(3)$ in the following as $M_{N}$.
(\ref{Laughlin}) exhibits a large overlap with the true, numerically
determined ground states for small $N$ in the CM.
This was shown only for up to four electrons on the disk \cite{Lau83b} and for
the corresponding wave function on the sphere for up to ten electrons
\cite{FOC86}. There are many arguments why (\ref{Laughlin}) is so successful.
In particular, if $q$ is not too large ($q \leq 70$) \cite{CLWH82}, it
describes an incompressible, homogeneous liquid \cite{Lau90a,CTZ93b}.
For a finite number of interacting electrons on the sphere, the requirement of
a
homogeneous and nondegenerate state makes the identification of special
spectra corresponding to filling factor $1/q$ simple,
because the ground state must have zero total angular momentum \cite{HR85a}.
\\
We want to identify -- in the disk geometry -- in a series of spectra for
the two interactions, CM and SRM, the ground state of the filling factor at
$1/3$ as a special state which we call the \underline{stable state} and which
we compare with the Laughlin wave function.
There will be problems in the case of the CM, but the identification will be
straightforward for the SRM. \\
We start with the CM (with background potential included) (\ref{H}).
For an overview, the complete spectrum of a system with 6 electrons on the disk
calculated by the exact diagonalization procedure is shown in Fig.\ 1a
ordered with respect to $M$.
We have chosen $m_{max}=17$ so that the naive filling factor is $\nu '=1/3$.
We use $h=m_{max}$ in (\ref{H}); numerical tests show that larger $h$ have very
small influence on the energy values but just increase the dimension of the
Hilbert space.
The ground state is nondegenerate and has $M_{6}=45$ as predicted by the
Laughlin function.
Still, it is not {\em a priori} clear that the ground state for this choice
of $m_{max}$ is a finite $N$ approximation of the $1/3$--state.
We will investigate this now by changing $m_{max}$; generally, as we increase
$m_{max}$ from its minimum value $N-1$ (at $\nu '=1$), the spectrum moves to
larger $M$ and the ground state angular momentum increases.
In order to show this in detail, we compare for 7 electrons the low--energy
levels in the vicinity of $\nu '=1/3$ where $m_{max}$ varies between 16 and 21
(Fig.\ 2a - f); the spectrum at $\nu '=1/3$ is given in Fig.\ 2e.
At first glance, the most promising candidate for a particularly stable ground
state seems to be the one in Fig.\ 2a, $m_{max}=16$, where the gap between
the ground state energy level and the first excited state is the largest.
But this $m_{max}$ does not correspond to $\nu '=1/3$.
Further, the state with one flux quantum more ($m_{max}=17$) does not show any
signature of a degeneracy in the ground state in contradiction to the \QP\
picture which is discussed in detail below in Section 5.
Finally, the ground state spectrum with the formally correct $\nu '=1/3$ given
in Fig.\ 2e does not exhibit the expected total angular momentum for seven
electrons $M_{7}=63$, but instead, $M=57$.
Table 1 summarizes the dilemma.
It gives, from 5 to 7 electrons, the total angular momentum $M_{g}$ of the
exact ground state for a range of filling factors around 1/3 expressed by
$N_{\Phi }=m_{max}+1$.
This range is chosen such that within it, the ground state angular mommentum
$M_{g}$ increases through its Laughlin value $M_{N}$.
Also, $m_{max}$ and the angular momentum of (\ref{Laughlin}) are given.
The table shows that for the CM and $N=5,6,7$, an identification of the
$1/3$--ground state cannot be brought into agreement with what the Laughlin
theory predicts.
Either the maximum single particle angular momentum $m_{max}$ or the total
angular mommentum $M_{g}$ does not agree with (\ref{Laughlin}).
Certainly, finite size corrections could in principle, in the limit
$N \rightarrow \infty$, remedy this problem.
But the tendency seems to be in the opposite direction.
The situation does not improve at all for eight electrons where the ground
state at $\nu'=1/3$ has an angular momentum 77 unequal to $M_{8}=84$. \\
Disregarding this difficulty, we extrapolate the ground state energy per
particle at $\nu '=1/3$ for the block with Laughlin's value $M=M_{N}$ from data
for up to 9 electrons to the thermodynamic limit.
In the case $\nu '=1$, the finite size corrections scale as
$\frac{1}{\sqrt{N}}$. Assuming this behavior also at $\nu' =1/3$ yields
\begin{equation}
\frac{E_{g}(\frac{1}{3})}{N}=-0.409510 + 0.036220 \frac{1}{\sqrt{N}} +
O(\frac{1}{N}) .
\label{Eg}
\end{equation}
Fig.\ 3 shows the exact data and the extrapolation.
Eq.\ (\ref{Eg}) is in good agreement with various other exact diagonalizations
and with energy expectation value calculations using Laughlin's trial wave
function on the disk \cite{MH86} and on the sphere \cite{FOC86,HR85a,MH87}. \\
We now change the model in two steps to arrive at the SRM and do a similar
study there.
At first, starting from the spectrum Fig.\ 1a, we omit the background.
This changes the shape of the spectrum dramatically, see Fig.\ 1b.
Without background, the repelling electrons tend to occupy the largest
available angular momenta. In such a case, the Laughlin wave function cannot
describe the ground state correctly.
In the second step, changing the interaction to the SRM, we calculate the
spectrum without background still using the same $m_{max}=17$.
Then, the ground state of this spectum is highly degenerate, see Fig.\ 1c.
This shows that a change from the CM with background to the SRM using the same
external parameters ($N, m_{max}$) does not necessarily yield a
nondegenerate ground state, as we expect at the stable state. \\
In order to find the nondegenerate ground state, we decrease $m_{max}$
successively and find for $N=6$ at $m_{max}=15$ really a nondegenerate state of
zero energy (Fig.\ 1e, for comparison see also the Figures 4a--f for 7
electrons including the stable state in Fig.\ 4c).
For completeness, we show in Fig.\ 1d also the spectrum for the case of the SRM
with a background where only $V_{1}$--parts contribute to both
electron--background and background--background interaction in (\ref{H}). \\
The nondegenerate ground state as seen in the spectra of Fig.\ 1e and 4c for 6
and 7 electrons, respectively, can be immediately identified as the Laughlin
wave function (\ref{Laughlin})  for $N=6$ and $N=7$.
It is well known that (\ref{Laughlin}) is the exact and only solution of
zero energy in the SRM at $\nu =1/3$ because it is the only $N$--electron
fermionic wave function not containing a component with relative angular
momentum one for any two electrons \cite{Hald83,PT85,TK85}.
The spectra display the expected ground state angular momentum $M= M_{N}$
(in Fig.\ 1e, $45=M_{6}$ and in Fig.\ 4c, $63=M_{7}$).
The situation is similar in the case of the sphere where the stable state has
$\vec{L}^{2}=0$ and is thus nondegenerate and homogeneous.
Eventually, this finding justifies our use of $\nu $ which differs from the
naive definition $\nu '$ by a finite size correction, since $\nu =1/3$ for the
parameters of Fig.\ 1e and 4c. \\
After having compared the spectra of the CM and the SRM at filling factor $1/3$
we conclude that the SRM seems to be -- on the disk -- the most promising model
for a discussion of the concept of a stable $1/q$--state and of \QPs.
This is true although thermodynamic instabilities are expected in this model
\cite{GM90}.
In contrast to the case of the spherical geometry, where the results for the CM
do not differ qualitatively from those for the SRM \cite{Hald90,HN86}, the SRM
suggests itself for a study in the disk geometry.
Thus, from now on, we will exclusively consider the SRM.  \\
As a side remark, we supply an argument supporting the nondegeneracy of
(\ref{Laughlin}) in the disk geometry.
We try to construct a state degenerate with (\ref{Laughlin}) by applying a
magnetic translation (commuting with $H$ for the infinite system)
to (\ref{Laughlin}) and show that this leads back to (\ref{Laughlin}).
\begin{equation}
\hat T_{\xi } \Psi _{\frac{1}{q}}(z_{1},\ldots,z_{N})=\Psi
_{\frac{1}{q}}(z_{1},\ldots,z_{N}) e^{-
\frac{N}{4}|\xi |^{2}}\prod_{i=1}^{N}e^{\frac{1}{2}z_{i}\xi ^{*}}
\label{TaufPsi}
\end{equation}
An expansion of the product w.r.t. $\xi ^{*}$ yields eigenfunctions with total
angular momentum $M_{N}(q)+l$
\begin{equation}
\sum_{l=0}^{\infty} \sum_{k=0}^{l} \sum_{j} {\cal N}_{l,k,j}(\xi ^{*} ) \; \;
P^{(l)}_{k,j}(z_{1},\ldots,z_{N})\Psi_{\frac{1}{q}}(z_{1},\ldots,z_{N}).
\label{TaufPsirhs}
\end{equation}
The $P^{(l)}_{k,j}$ are symmetric polynomials of degree $l$, $k$ counts the
maximum power of any variable $z_{i}$ ($0 \leq k \leq l $) occuring in
$P_{l,j}^{(k)}$ , and $j$ enumerates the individual polynomial for fixed $l$
and $k$.
If $\Psi _{\frac{1}{q}}$ is an exact eigenstate, as it is e.\ g.\  for the SRM,
then for an infinite large disk all the functions
$P^{(l)}_{k,j}\Psi _{\frac{1}{q}}$ are again degenerate eigenfunctions, but
with $\nu  =\frac{N-1}{q(N-1)+k}$.
However, we search for eigenfunctions with $\nu =1/q$.
Thus, we have to put $k=0$ which leaves in (\ref{TaufPsirhs}) only the term
$l=0$, the Laughlin state (\ref{Laughlin}).
This shows how a translation of (\ref{Laughlin}) followed by a restriction to
the orignal area of the system (filling factor) leaves (\ref{Laughlin})
invariant (see for the case of other topologies \cite{WN90}).
A similar procedure will prove useful in the case of the \QPs\, see below. \\
Finally, we comment on the degeneracy in the spectra of systems without
background potential for small $M$.
$M$--blocks between $M_{min}$ and $ M_{min}-(N-1)+m_{max}< M_{N}(q)$ are
entirely unaffected by the confinement (this is only true without background).
Energy levels with different $M$ are degenerate as the result of the
center of mass motion (in Fig.\ 1b and 1c the part of the spectra between
$M=15$ and $M=27$, and in Fig.\ 1e the part between $M=15$ and $M=25$).
In some sense, even the Laughlin wave function is also such a state because
enlargement of the system does not influence the energy of the state with zero
energy. \\
It should be mentioned that we have refrained from calculating the overlap
between Laughlin's wave function and the numerically calculated ground states
because the agreement in $M$, in connection with the energy of the state being
zero, suffices for our conclusion.
Summarizing our attempts to identify the $1/q$--Laughlin state in numerically
calculated spectra we see that for the SRM, the ground state can be identified
uniquely already for a finite system. In the CM case, this is not possible.
While on the sphere, the spectra of the two interactions for finite $N$ are
supposed to be related by perturbation theory, on the disk, they look
completely different.

\section{Quasiparticles}
After having identified the stable state at the filling factor 1/3 in the SRM,
we turn to the ground state properties at filling factors nearby.
Laughlin \cite{Lau83b} was the first who introduced the notion of \QPs\ at
$1/q$ in the disk geometry.
Here, we look for a definition applicable in our finite system.
Starting from the stable state, we create the two kinds of \QPs\ by increasing
and decreasing, respectively, the number of flux quanta through the area.
The \underline{quasihole} is then defined as the bulk ground state for a system
with one additional flux quantum through the area covered by the electrons,
i.\ e.\, $m_{max}=q(N-1)+1$, and a \underline{quasielectron} is the bulk
ground state for a system with one flux quantum less, i.\ e.\ ,
$m_{max}=q(N-1)-1$. This leads to a smaller and larger filling factor,
respectively, with a deviation of the order $1/N$ from $1/q$.\\
In general, the \QPs\ are fundamentally different from ordinary \QPs\ of a
Fermi liquid in that they cannot be constructed from the original fermions by
a process of switching on the interaction adiabatically.
They have various descriptions as anyons \cite{Halp83}, bosons \cite{Hald83}
or fermions \cite{Lau90a}.
They are not elementary particles as the electrons, but charge deviations from
the homogeneous density of the Laughlin state.
There are two applications of this notion of \QPs.
The first is the description of excited states at an unchanged filling factor
$1/q$.
In this picture, quasihole and quasielectron, separated by an infinite
distance,
form a quasiexciton \cite{Lau84a} which is a short--wavelength--excitation
($k \rightarrow \infty$) \cite{GMP86a,Hald85,KH84}.
The energy of this excitation is the sum of the two \QP\ energies; this is the
gap seen in the activation measurements of $\rho _{xx}$ \cite{WSTGE88}.
Otherwise, for small $k$, a collective theory yields the low lying
excitations \cite{GMP86a}. \\
Here, we are interested in the other use of the \QPs\ in which they describe
the ground state properties near $1/q$ and in which they are the elements of
the hierarchical theory \cite{Hald83}.
This theory sets out to understand the occurence of filling factors $p/q$
($p \not= 1, q$-- odd).
Starting with $N$ free electrons in $N_{\Phi }$ energetically degenerate
states in the lowest Landau level and switching on an interaction (e.\ g.\
the SRM--interaction), one gets a stable state for the special filling
factors $\nu = 1/q$, i.\ e.\ for $N_{\Phi }-1=q(N-1)$.
The hierarchical theory now draws an analogy from the electrons to the \QPs.
These can occupy of the order of $N$ degenerate states.
Creating a macroscopic number $N'$ of \QPs\, one gets in the presence of a
\QP--\QP\ interaction a new stable state, the "daughter state on the first
level of the hierarchy", for the special filling factor $1/p$ of the \QPs,
i.\ e.\ for $N = p(N'-1)$ ($p$ -- even for bosonic \QPs).
Then, the electronic filling factor is given by
\begin{equation}
\nu =\frac{N-1}{(N-1)q \pm N'}
\stackrel{N \rightarrow \infty}{ \longrightarrow}\frac{1}{q \pm
\frac{1}{p}} .
\end{equation}
For $q=3$ and $p=2$ this gives filling factors of $2/5$ and $2/7$.
Iterating the scheme, one can find successively all filling factors between 0
and 1 with odd denominators in an unique manner on different levels of the
hierarchy. \\
We want to emphasize that there are several propositions which the \QPs\ have
to fulfill for a justification of this theory. \\
-- The lowest one--\QP\ levels must be energetically degenerate at least
in the thermodynamic limit, and there must be an energy gap to the other states
of higher energy. \\
-- For a two--\QP\  system (two flux quanta more or less), it must be possible
to identify the low energy states as two free \QPs\ plus an interaction
contribution.
I.\ e.\ , these states must be separated from the energetically higher states
which means that the interaction must be small enough to leave a gap between
the two--\QP\ states and the neglected upper states. \\
-- The low--energy levels for more than two \QPs, particularly for a
macroscopic number of them, must be describable by an effective Hamiltonian
with up to two--particle interactions as determined from the two--\QP\ spectra.
Higher interactions are not allowed. This is the strongest condition and the
test for the validity of the hierarchical theory itself. \\
In summary, the low--energy behavior of the effective \QP\  Hamiltonian has to
reproduce the low--energy behavior of the original fermionic problem.
For the simplest non--trivial case of quasiholes at $\nu =1$, this has been
successfully checked on the disk with a special interaction \cite{Kas94}. \\
In the past, there were a lot of efforts to justify quantitatively the \QP\
picture.
Almost all authors used the spherical geometry and thus neglected the effect
of boundaries.
The work was mainly concentrated on the determination of the \QP\ energies at
various filling factors \cite{Lau83b,MH86} and then, conclusions were drawn
for the hierarchical theory.
B\'eran and Morf \cite{BM91} extracted the effective interaction of
quasielectrons on the sphere and estimated the ground state energy and the gap
of the $2/5$--stable state.
Endesfelder and Terzidis \cite{ET92} extrapolated the interaction of
quasiholes and compared the spectra of a small number of bosons interacting via
this interaction with electronic spectra of the same number of quasiholes.
There is a large asymmetry between quasiholes and quasielectrons \cite{ZX89}.
Their energies are of a different order of magnitude, and the subspaces of low
lying excitations have different dimensions for the quasiholes and the
quasielectrons.
Thus, it was tried to explain these different dimensions with a difference in
the \QP\--interaction: both \QPs\ are bosons, but quasielectrons have a
hard--core interaction. \cite{HXZ92}. \\
%
%
In the two following sections, we want to study whether a degeneracy of the
\QPs\
is found in the disk geometry. In the spherical geometry, this degeneracy
holds
trivially because the angular momentum of the \QPs\ is
$\vec{L}^{2}=\frac{N}{2}(\frac{N}{2}+1)$ and, therefore, there are $N+1$
degenerate states ($L^{z}$--degeneracy) \cite{Hald85}.
In contrast to the authors cited above, we investigate the \QPs\ in
the disk geometry, where the situation is more difficult as a result of the
boundary.

\section{Quasiholes}
A quasihole is created by adding one flux quantum to the system at $\nu = 1/q$
by increasing the area of the stable state system by one flux quantum, i.\ e.\
,
$m_{max}=(N-1)q+1$.
For the short--range model, we know already such a state, an exact
eigenfunction
of zero energy, i.\ e.\ put $k=1$ in (\ref{TaufPsirhs}).
Expanding the exponential in (\ref{TaufPsi}) up to first order in $z_{i}$ (and
substituting $\xi ^{*} \rightarrow  -2/\xi $) one finds a function of the form
\begin{equation}
\prod_{i=1}^{N} (z_{i}-\xi ) \Psi_{\frac{1}{q}}
= \sum _{l=0}^{N}  (-\xi )^{N-l} \: \Psi _{\frac{1}{q}}^{(-,M_{N}(q)+l)}
\label{loch}
\end{equation}
i.\ e.\ , the quasihole wave function earlier proposed by Laughlin
\cite{Lau83b}
as a good trial wave function for the CM.
The r.\ h.\ s.\ of (\ref{loch}) defines the expansion in components with
angular
momenta $M=M_{N}(q)+l$.
Thus, there are $N+1$ degenerate states with zero energy as on the sphere. \\
The inclusion of the Laughlin wave function (\ref{Laughlin}) in (\ref{loch}) as
the quasihole component $l=0$ is the consequence of a somewhat ambiguous
situation.
The quasihole has a local charge deviation from the homogeneous value
$\frac{\nu }{2\pi }$ of the charge distribution of the electrons on the disk.
E.\ g.\ for $\nu =1$, the one--particle density of a quasihole at $\xi $ is
\cite{MG86b}
\begin{equation}
n^{(-,\xi )}(z) = \frac{1}{2\pi }(1 - e^{- \frac{1}{2} |z-\xi |^{2} } ) \:.
\end{equation}
Around $\xi $ there is a charge depletion of magnitude $e^{*}=\nu e $.
For a finite system, the charge missing at $\xi $ accumulates at the finite
edge reflecting the enlargement of the disk area by one flux quantum.
The component of (\ref{loch}) with the largest possible total angular
momentum $M=M_{N}(q)+N$ is identical to the quasihole at $\xi =0$.
If the parameter $\xi $ is considered as a particle coordinate for a quasihole
in the lowest Landau level, smaller angular momenta of the electrons
correspond to larger radii for the quasiholes \cite{Lau90a}.
Then, the component of (\ref{loch}) with the smallest $M=M_{N}(q)$ corresponds
to the quasihole on the border of the disk and cannot be distinguished from the
Laughlin state.
In this sense, the Laughlin state (\ref{Laughlin}) is a quasihole too and
therefore, for the creation of a quasihole at a location $\xi \not= 0$, this is
also needed.  \\
The angular momentum component corresponding to a quasihole at $\xi =0$
can be chosen to generate the other quasihole components with the help of the
magnetic translation as follows.
Starting with $ \prod_{i=1}^{N}z_{i} \Psi_{\frac{1}{q}} \:,$
applying $\hat T_{\xi } $, and annihilating all total angular momentum
components with a filling factor corresponding to $m_{max} > (N-1)q+1$
one recovers the quasihole function (\ref{loch}).
Physically, this means a shift of the quasihole at $\xi =0$ against the
homogeneous background of the Laughlin wave function followed by a cut off at
the border of the original system. \\
Now let us turn to the spectra for one quasihole. For the Coulomb case, none of
the spectra for 7 electrons in Fig.\ 2 shows any evidence of a degeneracy
around
$\nu '=1/3$ in the low energy region. This observation expresses again the
difficulty with the interpretation of the spectra with Coulomb interaction for
a
finite number of electrons.
Therefore, coming back to the SRM, we discuss Fig.\ 4d (7 electrons).
Actually, the ground state is $(N+1)$--fold degenerate and has zero energy and
angular momenta reaching from $M=M_{N}$ to $M_{N}+N$ (here $M=63$ to 70).
This is obviously in agreement with the fact that the state (\ref{loch}) is a
zero energy eigenstate, which follows because the multiplication with powers of
$z_{i}$ can not change the minimum relative angular momentum from 3 in
Laughlin's wave function.
The low energy states, appearing in branches above the zero ground states with
excitation energies of an order of magnitude smaller than the energy scale
given
in the SRM by $V^{SRM}_{1} \simeq 0.44$, are connected with the edge
excitations
\cite{Sto91}. \\
We want to determine the quasihole energy in the SRM. Since there are three
external parameters in our model, three mechanisms exist for the creation of
\QPs. Correspondingly,
the three energies (the indices $\pm$ mean quasihole and quasielectron,
respectively) are the gross \QP\  energy $\varepsilon_{\pm} $ (change of the
electron number), the proper \QP\  energy  $\tilde{\varepsilon }_{\pm}$
(change of the magnetic field) and the neutral \QP\  energy
$\varepsilon ^{n}_{\pm}$ (change of the area) \cite{MH86,MG86b}.
There are relations between these energies involving the ground state energy
per
particle of the stable state. Thus, it is sufficient to know one of the
energies.
The energy taken from our spectra is the neutral \QP\  energy because
we change
the disk area for the creation of the \QP.
In the SRM model with zero ground state energy, all these energies are equal,
i.\ e.\ , $\varepsilon =\tilde{\varepsilon}=\varepsilon ^{n}$ ($\nu <1$).
The neutral quasihole energy for the $\nu=1/q$--state is defined as the
difference of the ground state energy of the one--quasihole spectrum
($m_{max}=q(N-1)+1$) and that of the stable state ($m_{max}=q(N-1)$).
We find from the exact spectra in agreement with the analytical arguments given
above $\varepsilon ^{n}_{-}(1/3)=\tilde{\varepsilon}_{-}(1/3)=
\varepsilon_{-} (1/3)=0$.\\
Two--quasihole spectra are pictured in Fig.\ 1c for 6 electrons and in Fig.\ 4e
for 7 electrons ($m_{max}=20$), where we have again a degenerate ground state.
But in contrast to the one--quasihole spectrum, where the zero energy levels in
one block (a line) were non--degenerate, the levels in this case are partially
degenerate; e.\ g.\ , in Fig.\ 4e we find for the $M$ reaching from 63 to 77
($M=M_{7}$ to $M=M_{7}+2\cdot 7=77$) energy degeneracies of
1, 1, 2, 2, 3, 3, 4, 4, 4, 3, 3, 2, 2, 1, 1.
This subspace of the lowest energy eigenstates (here of zero energy) can be
considered as being made up of bosonic quasiholes.
The quasihole can occupy $N+1$ one--particle states, thus $d$ bosons
(quasiholes) form a Hilbert space of dimension ${(N+1)+d-1\choose d} =
{N+d \choose d}$. In particular, the total degeneracy for two quasiholes is
just
${N+2 \choose 2}$, i.\ e.\ , for $N=7$, there are 36 states, and the exact
spectrum actually shows this. In general, the degeneracy $g_{B}(N,d,\tilde{M})$
of a level with $M=M_{N}(q)+\tilde{M}$ in a $d$--particle Bose system with
$N+1$
one--particle states can be derived from the following generating function
which
is analogous to the fermionic formula (\ref{gF}):
\begin{equation}
\prod_{l=1}^{d} \frac{1-x^{(N+1)+(l-1)}}{1-
x^{l}}=\sum_{\tilde{M}=0}^{Nd}g_{B}(N,d,\tilde{M}) x^{\tilde{M}}.
\end{equation}
Additionally, we confirm the validity of this description by inspection of the
degeneracy of the zero energy levels of, e.\ g.\ , the three--quasihole
spectrum
in Fig.\ 4f ($d=3,N+1=8, \tilde{M}=0,\ldots,21$). On the other hand, it is
immediately clear that the total dimension of the Hilbert space is actually
$\sum_{\tilde{M}=0}^{Nd}g_{B}(N,d,\tilde{M}) = {N+d  \choose d}$.  \\
We summarize that the 1/3--quasihole spectra in the SRM corraborate strongly
the quasihole picture already for finite $N$.
Unfortunately, the interaction between the quasiholes is zero.
A bosonic description of the quasiholes appears quite natural considering the
degeneracies in our spectra.
This conclusion is in agreement with investigations on the sphere where the
Hilbert--space dimension can be understood in the boson--picture
\cite{Hald83,HXZ92}.

\section{Quasielectrons}
In the hierarchical construction, the \QPs\ enter symmetrically, but in
their energies and spectra, they are not very symmetric \cite{ZX89}.
For the energies, this is quite inevitable, since a 1/3--quasielectron is
created by decreasing the maximum angular momentum to $m_{max}=q(N-1)-1$
in order to reach a larger $\nu $.
Confining the electrons to a smaller disk gives rise to a stronger
interaction contribution compared to the case of an enlargement of the disk.
Thus, $\varepsilon^{n}_{+}(1/3)$ must be positive. \\
Since the SRM plays the role of a canonical model, particularly with respect
to the quasiholes, we hope that it reflects the essential physics in the
quasielectron case too. Still, the situation for the quasielectrons is
different from that for the quasiholes because we can not expect zero
energy eigenvalues and we do not know an exact eigenfunction for the
quasielectron up to now. \\
Let us look at the one--quasielectron spectrum for the SRM for 7 electrons in
Fig.\ 4b.
The ground state at $M_{7}=63$ has a non--zero energy.
This is to be expected because it is not possible to avoid relative angular
momentum one in any wave function with $M \leq M_{N}$ and $m_{max}=3(N-1)-1$.
Further, there are low lying energy levels with $M \geq M_{N}$ (see Fig.\ 4b)
which are, as in the case of the quasihole, again attributed to the edge of
the system.
The quasielectrons are the lowest energy levels with $M=M_{N}-N,\ldots,M_{N}-1$
(here from M=56 to 62) whose energies are almost degenerate. \\
While this seems to be in contradiction to the usual definition of the
quasielectron as a ground--state of the spectrum, we justify our focus
on these angular momentum blocks by considering the quasielectron trial wave
functions.
That there are many proposals for such trial wave functions
\cite{Lau83b,MH86,MG86b,KA92,Jai89b} reflects the fact that one has for the
quasielectron a more complicated situation than for the quasihole.
Still, all proposals have angular momentum components in the range of
$M=M_{N}-N,\ldots,M_{N}$.
This becomes particularly clear, if on considers Haldane's quasielectron on
the sphere \cite{Hald83}.
This $N+1$--fold degenerate wave function is a good trial wave function
for the description of the quasielectron which is well exhibited in the
spectra on the sphere \cite{Hald83,HXZ92}.
After a projection onto the plane according to the rules given in \cite{FOC86}
the resulting wave function has $N+1$ angular momenta components reaching from
$M=M_{N}-N$ to $M_{N}$ and this is our trial function introduced in
\cite{KA92}.
The component with $M=M_{N}$ has much smaller energy than the other components
and is therefore excluded. \\
After having identified the angular momenta of the quasielectron components
we want to evaluate the energies.
The exact energies for up to 8 electrons have been compared with those of
Laughlin's, Jain's, MacDonald's and our quasielectron trial wave functions
in our previous work \cite{KA93a}.
We extend here this comparison to 9 electrons.
The exact energies for $N=9$ are pictured in Fig.\ 5a together with the trial
function data from Table 1b in \cite{KA93a};
the data from Table 1c in \cite{KA93a} for $N=10$ (without results from exact
diagonalizations) are shown in Fig.\ 5b.
For both particle numbers, the pictures are very similar with respect to the
$M$--dependence of all trial wavefunctions; the exact eigenenergies are
almost independent of $M$.
Jain's quasielectron wave function has, in comparison with the other trial
functions, the best energy expectation values, but for higher $M$, the
deviations from the exact values are still quite large.
The exact data in the one--quasielectron spectrum exhibit already for a finite
electron number almost degeneracy.
This degeneracy is not so well expressed in Fig.\ 4b as in Fig.\ 5a, because
of the different scale; the relative deviation from the average value is for
$N=9$ only 5.6 \%.  \\
Next we try to find the neutral quasielectron energy
$\varepsilon ^{n}_{+}(1/3)$ from the exact numerical data for up $N=9$
electrons in the SRM.
Unfortunately, an extrapolation with respect to $N$ for a single component of
the quasielectron, e.\ g.\ for the one with the smallest angular momentum
$M=M_{N}-N$, does not work because the energy fluctuates with $N$.
Therefore, we calculated the arithmetic mean of the quasielectron
energies of all $N$ components forming the new quantity
$<\varepsilon ^{n}_{+}(1/3)>$.
Then, these data are extrapolated to the thermodynamic limit by minimizing the
weighted quadratic deviation from an extrapolation curve with a finite size
correction proportional to a negative rational power of $N$ to be determined.
This way, we find in the thermodynamic limit
\begin{equation}
<\varepsilon ^{n}_{+}(1/3)> = 0.1865 + 1.8790 \; N^{-2.6320}
\label{epsplus}
\end{equation}
This curve is shown in Fig.\ 6a.
The finite size corrections vanish surprisingly fast.
Arguments given by Morf et al.\ \cite{MH86,DM89} for a $N^{-\frac{3}{2}}$
correction are not applicable here because there is no background in the SRM.
The result (\ref{epsplus}) can be compared with the numerical data for the
discontinuity in the chemical potential in the case of short--range interacting
electrons on the surface of the sphere.
We extrapolate from Table 2 in \cite{GM90} the value for the jump of the
chemical potential and find that the
neutral quasielectron energy at $\nu =1/3$ is 0.1905 showing quite good
agreement between our extrapolation and the result of the other geometry.  \\
We now extrapolate the energy expectation values of the trial wave functions,
see Fig.\ 6b, again using the average of the available components at a given
$N$.
Here, the leading finite size correction is unknown too.
The extrapolation of Jain's quasielectron is closest to the exact data.
On the other hand, the finite size correction is qualitatively different from
that of the exact data, see Fig.\ 6a, particularly the approximate
$1/\sqrt{N}$--behavior of Jain's trial wave function is not found in the exact
data.
Still Jain's proposal comes closest to the extrapolated value of 0.1865.
The proposal of \cite{KA92}, on the other hand, yields a higher limit, but the
behavior of its finite size correction is much more similar to the exact
result. \\
One of the most interesting questions for a test of the relevance of the
notion of a quasielectron is the understanding of the many--\QP\ behavior.
The simplest case is the two--quasielectron spectrum shown for 7 electrons for
the SRM (Fig.\ 4a).
We realize that different from the case of the two--quasihole spectrum
there is a complicated interaction superposed by finite size effects.
It seems difficult to identify the two--particle states arising from
the subspace generated by the quasielectrons. One of the reasons should be a
strong interaction between the quasielectrons leading to a mixing with other
energy levels of the many--electron system.
Thus, a detailed investigation of the quasielectron interaction for two
quasielectrons and further implications for more quasielectrons in the
framework
of the hierarchical scheme remain a matter of future research.
Now, there is an intensive discussion on this point
\cite{DJ92b,HXZ92,YS93} and a satisfying explanation of the success of Jain's
wave function and its connection with the ideas of the hierarchical scheme
are still missing. We do not believe that a proposal reconciling the
two pictures \cite{YS93} is justified up to now.

\section{Summary}
The numerical diagonalization of the FQHE Hamiltonian for up to nine electrons
on the disk shows that whether a unique identification of a particular, stable
state at filling factor $1/q$ is possible in the spectra depends on the
electronic interaction.
This is in contrast to the spherical geometry.
Whereas for the Coulomb interaction, the ground states of spectra around $1/3$
can not be related to the Laughlin wave function for small $N$, the latter
turns out to be the exact ground state solution of the model with short--range
interaction.
The picture that the spectra for Coulomb and for the short--range interaction
are very similar and that the states of both interactions are related
to each other by only a small perturbation could not be quantitatively
confirmed
for the present geometry of a finite disk. \\
For the identification of the stable state and the quasiholes, the SRM
serves as a canonical model.
Then, $N+1$ zero--energy levels can be exactly identified in the one--quasihole
spectra as quasihole components of the Laughlin type.
The degeneracy of these levels in the spectra for two and more quasiholes
justifies to treat the quasiholes as bosons. \\
In the quasielectron case, $N$ components of the quasielectron are identified
which show almost degeneracy with small deviations from the average value in
our
finite size calculations.
The expectations values of trial wave functions of various authors are compared
with the exact data.
They deviate with increasing angular momentum more and more from the exact
values leaving space for improvement.
An extrapolation of the average quasielectron energy to the thermodynamic limit
is in agreement with the results of other geometries.  \\
It should be interesting to use the \QP\ identification on the disk in order to
extend the present treatment to the quasielectron--quasielectron interaction.
Creation energies and interaction parameters could then be used as input
parameters in an attempt to formulate an effective theory for the
quasielectron.

\section{Acknowledgements}
We especially want to thank B.\ Kramer and to acknowledge gratefully numerous
discussions with the members of the theory group 8.1 at the
Physikalisch--Technische Bundesanstalt in Braunschweig as well as
with M.\ Hellmund, J.\ K.\ Jain, A.\ H.\ MacDonald, R.\ Morf and W.\ Weller.

\newpage
\noindent   \underline{Table 1} \\
\begin{tabular}[t]{|c|c|c|l|}
\hline
 N & $N_{\Phi }=m_{max}+1$ & $M_{g}$  & Remarks    \\   \hline
 5 &    &    &   \\
   & 13 & 25 & $\nu =1/3$       \\
   & 14 & 25 & $\nu =1/3$ + one quasihole                \\
   & 15 & 30 & $\nu =1/3$ + two quasiholes, $\nu '=1/3$              \\
   & 16 & 30 & ...                        \\
   & 17 & 30 & ...                                                \\
   & 18 & 30 & ...                                                \\
   & 19 & 35 & ...                                                \\  \hline
   & 13 & 30 & Laughlin state                     \\    \hline \hline
 6 &    &    &   \\
   & 16 & 35 & $\nu =1/3$        \\
   & 17 & 39 & $\nu =1/3$ + one quasihole                                 \\
   & 18 & 45 & $\nu =1/3$ + two quasiholes, $\nu '=1/3$          \\
   & 19 & 45 & ...                                                  \\
   & 20 & 45 & ...                                               \\
   & 21 & 45 & ...                                               \\
   & 22 & 51 & ...                                               \\  \hline
   & 16 & 45 & Laughlin state    \\   \hline \hline
 7 &    &    &                                                 \\
   & 19 & 51 & $\nu =1/3$                                       \\
   & 20 & 57 & $\nu =1/3$ + one quasihole                                 \\
   & 21 & 57 & $\nu =1/3$ + two quasiholes, $\nu '=1/3$                 \\
   & 22 & 63 & ...                                  \\
   & 23 & 63 & ...      \\
   & 24 & 63 & ...                                                \\
   & 25 & 69 & ...                                         \\ \hline
   & 19 & 63 & Laughlin state      \\    \hline  \hline
\end{tabular} \vspace{0.5cm}

\noindent   "Search for the stable state" in the Coulomb model (CM)

\newpage
\noindent   \underline{Figure Captions} \\[2ex]
Fig.\ 1: Complete many--particle energy spectrum for 6 electrons and
various interactions.
The levels are ordered with respect to the total angular momentum $M$.\\
a: Coulomb interaction with background, degeneracy 18
($m_{max}=17$), i.\ e.\ $\nu '=1/3$. \\
b: as a, but without background. \\
c: as a, but for the short--range interaction (\ref{SRM}) without background.\\
d: as c, but with background.\\
e: short--range interaction without background, $m_{max}=15$, i.\ e.\
$\nu =1/3$.  \\[2ex]
Fig.\ 2: Series of spectra showing for decreasing filling factor the low lying
states for 7 electrons with Coulomb interaction; $m_{max}=16,\ldots,21$
(Fig.\ 2a -- f).  \\[2ex]
Fig.\ 3: Energy per particle for the model with Coulomb interaction and
background at $\nu' =1/3$ in dependence on the number of electrons $N$ up to
$N=9$. The line extrapolates to the thermodynamic limit. \\[2ex]
Fig.\ 4: Series of spectra showing for decreasing filling factor the low lying
states for 7 electrons with the SRM interaction (\ref{SRM});
$m_{max}=16,\ldots,21$ (Fig.\ 4a -- f).
Fig.\ 4c shows the stable state, Fig.\ 4d and 4b the one--quasihole state and
the one--quasielectron state, respectively. \\[2ex]
Fig.\ 5: Numerically calculated expectation values for Laughlin's,
our \cite{KA92}, and Jain's quasielectron trial wave functions
(from top to bottom) arranged with increasing $M$;  \\
a:  $N=9$ electrons, the lowest levels (dottet line) are the exact energies.\\
b:  $N=10$ electrons.  \\[2ex]
Fig.\ 6: Averaged quasielectron energy in the SRM (\ref{SRM}) extrapolated to
the thermodynamic limit. \\
a: exact energies for up to $N=9$. \\
b: Jain's ({\bf x}), Laughlin's ({\bf +}) and our (${\bf \ast}$) \cite{KA92}
quasielectron trial wave functions for up to $N=10$.

\newpage


\end{document}